# Where have all the bulges gone?

Leonel Gutierrez[1], Antti Tamm[*,2], John Beckman[1], Louis Abramson[3], Peter Erwin[4], Mélanie Guittet[5]

[1] Instituto de Astrofisica de Canarias; [2] Tartu Observatoorium; [3] Columbia University; [4] Max-Planck Institut für extraterrestrische Physik; [5] Observatoire de Paris

**ABSTRACT.** Several recent studies indicate that bulges are more complex than merely structureless relaxed stellar systems. We study the HST images of a sample of 130 nearby early type (S0-Sab) disc galaxies and detect pure structureless bulges with the Sérsic index $n \geq 2$ for only 12% of the galaxies. Other galaxies show varied substructure in their inner regions (inner bars, inner spiral arms, inner rings) and sometimes contain no bulge at all. Inner substructure is more common for these galaxies, which also display structure at larger scales.

**MOTIVATION: How many galaxies do have a "classical" bulge?**

The high angular resolution imaging available today has brought us to revise the concept of classical bulges of disc galaxies. The nature and structure of bulges of nearby disc galaxies has been addressed in several recent papers. As a major topic, the disc-like structure and origin of many bulges – therefore called "pseudobulges" – is discussed (as reviewed by Kormendy & Kennicutt 2004). Some studies have demonstrated the existence of secondary bars, inner discs and nuclear rings inside barred galaxies (e.g. Erwin & Sparke 2002); also correlations between "pseudobulges" and their Sérsic indices have been studied recently (Fisher & Drory 2008).

Spheroidal components of galaxies host a substantial fraction of the stellar mass in the universe; however, no comprehensive statistical study of the structure of nearby bulges has been conducted so far, despite the wealth of publicly available high-level observational data. We start filling in the gap by asking ourselves, how many of the nearby bulges actually do contain additional structures and with how many pure, "classical" de Vaucouleurs' $R^{1/4}$-bulges are there in the local universe.

**SAMPLE: 130 early type disc galaxies**

Our sample has been constructed to be a maximally neutral representation of local early-type disk galaxies for which HST imaging is available. On the basis of the Third Reference Catalogue of Bright Galaxies (de Vaucouleurs et al. 1991), galaxies with Hubble types between S(B)0 and S(B)ab (Hubble stages T = -3.5 to 2) were selected as galaxies with the most prominent bulges by definition. Approximate distance limits were set for the sample; galaxies nearer than 10 Mpc were excluded because their bulges would not fit into the field-of-view of HST, while galaxies more distant than 50 Mpc were excluded because the spatial resolution of their imaging becomes too low. Also, galaxies at inclination angles higher than 60° were neglected because some of their bulge morphology might remain hidden. A total sample of 130 galaxies was formed, containing galaxies observed through the F625W, F775W, F814W or F850LP filters of the ACS camera or the F675W, F702W, F791W or F814W filters of the WFPC2 camera of the HST.

In many cases, the selected galaxies fill the field of view of HST. In order to have an idea of the extent and nature of the outer galactic structures, additional ground-based imaging was acquired from the Sloan Digital Sky Survey (SDSS) archive and the NASA/IPAC Extragalactic Database (NED).

**RESULTS: "classical" bulges are rare!**

According to structure detection or non-detection at different levels, the studied galaxies were categorized into three types: galaxies with classical bulges – no additional structure is detected at any level; galaxies with non-classical bulges – additional structures in the bulge region (bars, spiral arms, signatures of dust, star-forming regions) are clearly seen in the images; intermediate bulges – galaxies with faint structures in the bulge region, sometimes revealed only after unsharp masking or in residual images. The two latter types are possible indications of pseudobulges or missing bulges; however, each case needs a separate photometric and kinematic analysis for a clear-cut confirmation of this and we avoid referring to them as pseudobulges here.

---

[*] e-mail: *atamm@aai.ee*

Only 12 % of all the 130 galaxies have bulges without notable structure, while about two thirds (63 %) of the sample galaxies reveal easily detectable structural features in the inner regions, which are usually associated with disk morphology. Several representatives of the latter group present no obvious bulge-like spheroidal at all. The remainder are galaxies with generally classical bulges, but faint structures are revealed with unsharp masking or in the residual images.

Classical, featureless bulges become especially rare among Sa and Sab galaxies: our sample contains no such species – all Sa and Sab galaxies reveal at least faint structure. In general, galaxies possessing spiral arms, bars and rings in their general morphology also possess such features at smaller scales in their central regions. More details on this finding will be given in a forthcoming paper.

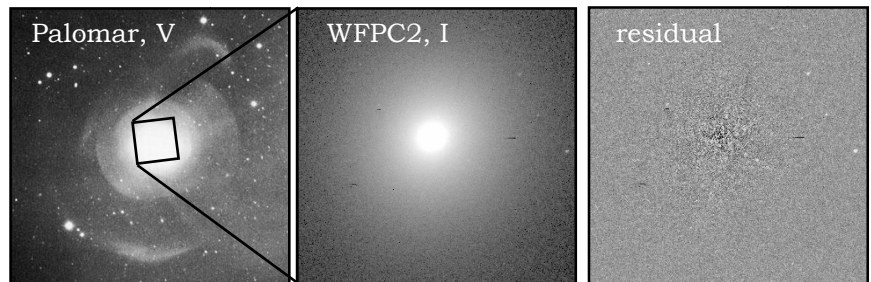

**Figure 1.** Galaxy NGC 0474, Hubble type S0. From left to right: ground-based imaging; the HST view; residual of the HST image after the subtraction of the azimuthally averaged model; azimuthally averaged surface brightness distribution. Despite the disturbed outer morphology, this galaxy hosts a "classical" featureless bulge.

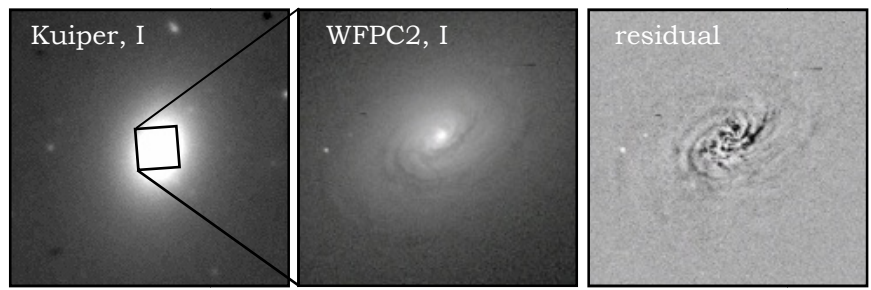

**Figure 2.** Galaxy NGC 2110, Hubble type S0. From left to right: ground-based imaging; the HST view; residual of the HST image after the subtraction of the azimuthally averaged model; azimuthally averaged surface brightness distribution. Clear spiral structure is present in the central region – this is a "non-classical" bulge.

**CONCLUSIONS.** Galaxies displaying structure (spiral arms, bars, rings) in the outer regions are very likely to possess structure also in the inner (bulge) regions. Pure, relaxed, featureless bulges are rare among disc galaxies. The results lay constraints on the formation scenarios of both bulges and discs and the whole cosmological framework. For further interpretation and confirmation of the results, 2D velocity fields have to be studied.